\documentclass[fleqn,10pt]{wlscirep}
\usepackage[utf8]{inputenc}
\usepackage[T1]{fontenc}

\usepackage{enumitem}
\usepackage{graphicx}
\usepackage{dcolumn}
\usepackage{bm}

\usepackage{xcolor}

\title{Mutation induced infection waves in diseases like COVID-19}

\author[1,*]{Fabian Jan Schwarzendahl}

\author[1]{Jens Grauer}%

\author[2]{Benno Liebchen}%

\author[1]{Hartmut L\"owen}%

\affil[1]{Institut f\"ur Theoretische Physik II: Weiche Materie, Heinrich-Heine-Universit\"at D\"usseldorf, 40225 D\"usseldorf, Germany
}

\affil[2]{Institute of Condensed Matter Physics, Technische Universit\"at Darmstadt, Darmstadt, Germany}

\affil[*]{Fabian.Schwarzendahl@hhu.de}

\date{\today}

\begin{abstract}
After more than 6 million deaths worldwide, the ongoing vaccination to conquer the COVID-19 disease is now competing with the emergence of increasingly contagious mutations, repeatedly supplanting earlier strains. 
Following the near-absence of historical examples of the long-time evolution of infectious diseases under similar circumstances, 
models are crucial to exemplify possible scenarios. 
Accordingly, in the present work 
we systematically generalize the popular susceptible-infected-recovered model to account for mutations leading to repeatedly occurring new strains, which we coarse grain based on tools from statistical mechanics to derive a model predicting the most likely outcomes. 
The model predicts that mutations can induce a super-exponential growth of infection numbers at early times, which self-amplify to giant infection waves which are caused by a positive feedback loop between infection numbers and mutations and lead to a simultaneous infection of the majority of the population. 
At later stages -- if vaccination progresses too slowly -- mutations can interrupt an ongoing decrease of infection numbers and can cause 
infection revivals which occur as single waves or even as whole wave trains featuring alternative periods of decreasing and increasing 
infection numbers. 
This panorama of possible mutation-induced scenarios should be tested in more detailed models to explore their concrete significance for specific infectious diseases.
Further, our results might be useful for discussions regarding the importance of a release of vaccine-patents to reduce the risk of mutation-induced infection revivals but also to coordinate the release of measures following a downwards trend of infection numbers.
\end{abstract}
\begin{document}
\flushbottom

\maketitle
\thispagestyle{empty}

\noindent\large \textbf{Introduction} \\ \normalsize
The COVID-19 pandemic~\cite{zhou2020pneumonia,wu2020new} has led to  more than 500 million infected~\cite{Dong2020} and more than 6 million death~\cite{Dong2020} worldwide until the beginning of Mai 2022.
During  the course of the pandemic the SARS-CoV-2 virus has mutated into various different strains~\cite{hadfield2018nextstrain,rambaut2020dynamic}, some of which have led to an increased infection rate~\cite{yao2020patient,korber2020tracking,grubaugh2020making} as compared to the original strain~\cite{wu2020new} (Wuhan 2019). 
Examples are the variants B.1.1.7 and B.1.351, which have driven a strong rise of infection numbers in the United Kingdom and South Africa~\cite{tegally2020emergence,priesemann2021action} in late 2020~\cite{volz2021assessing,davies2021estimated} and the P.1 mutation which has induced an infection wave in  Brazil~\cite{coutinho2021model} in early 2021. 
%Furthermore, both the variant B.1.1.7 and P.1 have been shown to have a higher mortality~\cite{faria_genomics_2021,davies2021increased}.
%New variants of the virus can spread faster, however, mutants may also reduce the effectiveness of antibodies  and vaccines~\cite{planas2021sensitivity,wang2021antibody,hoffmann2021sars,collier2021sensitivity,cele2021escape}. 
%Although new strains of the SARS-CoV-2 virus are quickly detected~\cite{hadfield2018nextstrain,rambaut2020dynamic}, and the genetic information is shared, identifying their transmissibility, mortality and relevance for the pandemic is usually only recognized once a strain is already wide spread.

The availability and ongoing vaccine production gives hope to slowly gain control of the disease~\cite{grauer2020strategic,zhouoptimizing,molla2021adaptive}. However, 
before herd immunity (if at all achievable) is finally reached worldwide it will take many month or even years, which the virus will exploit to mutate 
into a range of new strains. 
Thus, at the timescale of months or years a race is looming ahead between the occurrence of new mutations and the adaption and mass-production of existing vaccines to get these mutations under control.
In particular, this makes it questionable if the present (and future) vaccination programs are sufficiently effective to ultimately get diseases like COVID-19 under control.
%-- making it questionable if the present vaccination program is sufficiently effective to ultimately get COVID-19 under control. 
It is therefore important to understand possible mutation-induced long-time disease-evolution scenarios e.g. in view of the ongoing discussions regarding the release of patents to accelerate worldwide vaccination but also regarding requirement of measures like social distancing once the infection numbers show a downwards trend.

\begin{figure}[!t]
    \centering
    \includegraphics[width=0.7\columnwidth]{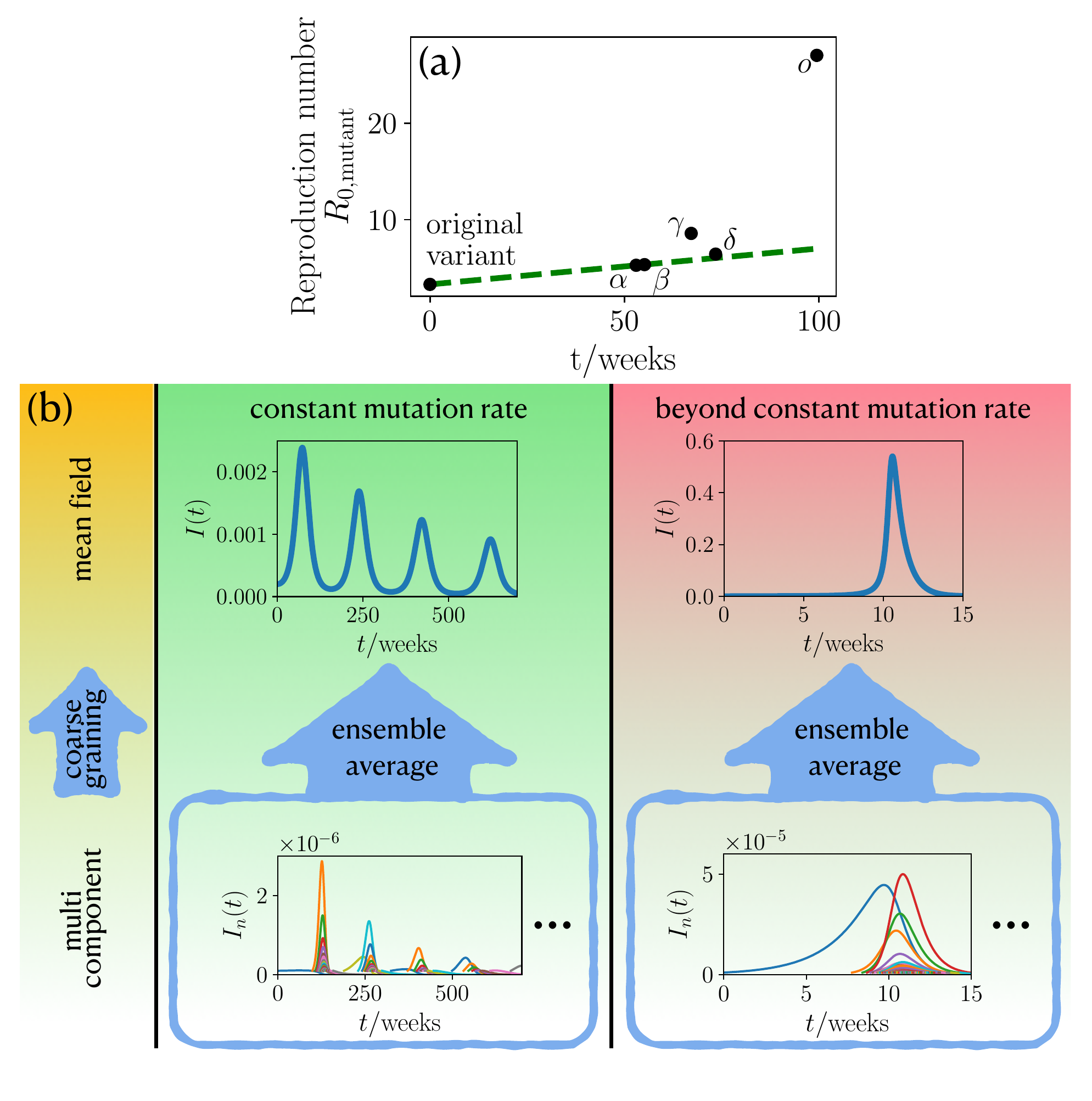}
    \caption{The statistics of mutation formation and its impact on the course of an epidemic.
    (a): Reproduction number for different COVID-19 mutations as a function of the their emergence time. (for details see Methods) Green dashed line shows a linear fit to initial constant growth.
    (b): For a constant mutation rate (middle panel, green background) an ensemble average of the multi component description with many infection strains $I_n(t)$
    leads to multiple infection waves of the global infection number $I(t)$.
    Beyond the constant mutation rate, if the mutation rate is coupled to the infection number (right panel, red background), 
    the ensemble average produces a hidden singularity, as manifested by the giant infection wave.   
    Only one representative realization of the multi component description is shown in the blue frames.
    The left panel (yellow background) indicates the different levels of description starting from multi components leading to an effective mean field by coarse graining.
    }
    \label{fig:model_sketch}
\end{figure}

Notably, historical examples to assess possible long-time consequences of mutation cascades are scarce, since particularly severe mutations have traditionally led to a rapid death of infected individuals eliminating these mutations. Thanks to modern medical treatment based e.g. on extracorporeal membrane oxygenation support or artificial aspiration, however, such a self-elimination of severe mutations is largely absent. Notably, besides the positive effect of immediately saving many lives, these treatments also have the side effect of inducing a potentially disastrous self-amplification of mutations and infection-rates. 
%This brings us to the assumptions that (i) mutations can serve as seeds for further mutations which are even more infectious than the strain from which they have emerged; (ii) mutation rates are higher when infection numbers are high. 
Here we are interested in particular in the effects of mutations on the spreading of an infectious disease in phases where (i) mutations can serve as seeds for further mutations some of which are even more infectious than the strain from which they have emerged and 
(ii) mutation rates are either constant or higher when infection numbers are high.
Both factors together can generally lead to a positive feedback loop between infection numbers and mutations suggesting severe long-time mutation-induced effects for the disease evolution. 
%Actual data for COVID-19 mutations show, in fact, early signatures supporting such a possible self-amplification scenario: 
Actual data for COVID-19 mutations show, in fact, early signatures supporting such a possible self-amplification scenario during some phases of the disease:
They reveal an initial constant and a subsequent nonlinear growth of the relevant infection rate
(Fig.~\ref{fig:model_sketch}(a)). 
%Thus, it would be highly important to understand the possible long-time consequences of such a self-amplification mechanism and how fast vaccination has to progress worldwide in order to suppress the most dramatic
For future pandemics, it would be highly important to understand the possible long-time consequences of such a self-amplification mechanism and how fast vaccination has to progress worldwide in order to suppress the most dramatic
ones. However, following the scarcity of useful historical examples illustrating the possible consequences, 
we have to rely on models to explore the possible impact of mutations on the long-time evolution of the disease dynamics, in particular also in the presence of vaccination and other actions counteracting the self-amplification mechanism. 
To provide a concrete starting point for such an exploration, in the present work, we develop a statistical minimal model to predict possible mutation-induced effects for the long-time evolution of infectious diseases like COVID-19. We first develop a stochastic multi-strain generalization of the popular susceptible-infected-recovered (SIR) model to  account for the random occurrence of mutations and then use the coarse-graining concept of statistical physics to derive an effective mean-field model enabling general predictions of the most likely scenarios for a given scenario (characterized by parameters such as the mutation and the vaccination rate). See Fig.~\ref{fig:model_sketch} (b) for an schematic illustration of our approach.

One generic prediction of our model is that mutations induce an \emph{explosive super-exponential growth} of the infection numbers rather than the ordinary and much discussed normal exponential growth, in phases where the population is far away from herd immunity. 
At later phases, when a population comes close enough to herd-immunity that the reproduction number drops below one ($R<1$) and infection numbers subsequently decrease to a very low level, mutations can raise the reproduction number to $R>1$ inducing a new infection wave, which is followed by a whole train of further waves. 
This scenario occurs even for a constant mutation rate (Fig.~\ref{fig:model_sketch} (b)). 
If the mutation rate increases with the number of infections, as generally expected and discussed above, their effect is even more dramatic: then, mutations occur at a self-accelerating pace and continuously prevent the population from reaching herd-immunity by persistently enhancing the effective reproduction number of the disease. As a 
result the infection dynamics approaches a hidden singularity and displays signatures of a critical dynamics. That is, infection numbers grow extremely fast, giving a giant infection wave, such that the majority of the population is infected at the same time (see the values on the vertical axis in Fig.~\ref{fig:model_sketch} (b)), which would massively overstrain any existing medical system.
Finally, in phases where vaccination of the population takes place and is sufficiently effective to suppress the hidden singularity and hence the explosive self-acceleration of infection numbers, our model predicts the possibility of mutation-induced infection wave trains, as in the case of constant vaccination, illustrating once more the possible dramatic consequences following from the fact that 
herd-immunity is not necessarily a permanent state in the presence of mutations. 
To see how these predictions come about, let us now discuss our general modelling approach in detail. Based on this approach we will then discuss our results for a constant mutation rate model and a model that goes beyond a constant mutation rate.

%{\color{blue} This mechanism entails an nonlinear growth of the infection rate of the virus. On the other hand for a constant mutation rate, the infection rate of the virus constantly grows. Mutations of the COVID-19 virus show an increase in the basic reproduction number and thus a similar growth in the infection rate (see Fig.\ref{fig:model_sketch}(b)). The rising infection rate favors an exponential growth and hence a mutation statistic that is coupled to the actual number of infected people.}

\vspace{1cm}
\noindent\large \textbf{Model} \\ \normalsize
To describe the impact of mutations on the infection dynamics within a simple statistical framework, we first generalize 
the popular susceptible-infected-recovered (SIR) model~\cite{kermack1927contribution,hethcote2000mathematics,andersson2012stochastic,grassly2008mathematical,Gog17209,harko2014exact,kroger2020analytical,schlickeiser2021analytical,bittihn2020stochastic,das2021scaling,yaari2013modelling,zhao2021contagion,norambuena2020understanding,day2006insights,day2007applying,day2004general,koelle2011dimensionless,koelle2010two,he2013inferring,boni2004influenza,levin1981selection}, which has been intensively explored in the context of the COVID-19 pandemic~\cite{maier2020effective,dehning2020inferring,te2020effects,duran2021more,te2021containing,lasser2021assessing,desvars2020structured,bittihn2021local,zhang2021epidemic,contreras2021challenges,estrada2020covid,yang2020mathematical}. 
While some recent works have generalized this model to account for two different infectious strains~\cite{gonzalez2021impact,fudolig2020local}, here we
allow for the continuous emergence of new strains with a rate $\nu$, which in general depends on the present infection number. 
Denoting the fraction of susceptible and recovered individuals with $S$ and $R$ respectively and the fraction of individuals which is infected with strain $n$ as $I_n$, this leads use to the following dynamical equations: 
\begin{align}
\dot{S}&=-\sum_n \beta_n S I_n, \label{eq:S_nmodel}\\
\dot{I}_n&=\beta_n S I_n-\gamma I_n, \label{eq:I_nmodel}\\
\dot{R}&=\sum_n\gamma I_n. \label{eq:R_nmodel}
\end{align}
Here, $\gamma$ is the inverse of the average disease duration, i.e. the recovery/death rate and $\beta_n$ is the infection rate of strain $n$, which we randomly choose from a certain characteristic distribution. 
As an initial state, we assume that initially (time $t=0$) we have only a single infectious strain with a low positive infection number such that only $I_0\gtrsim 0$ whereas $I_{n\neq 0}=0$ for $n=1,2..$. %Note that we do not include recovery rates or escape of immunity into our model.
%A representative infection pattern has many strains, shows super-exponential growth and can have multiple waves infection waves (see Fig.~\ref{fig:model_sketch} bottom).
\\To allow predicting the average (or most likely) result of the infection dynamics we now coarse grain this model, essentially by averaging over many strains and disease-realizations (see Methods section for technical details), which leads to the following effective model:
\begin{align}
\dot{S}&=-\beta(t,I) S I, \label{eq:S_cmodel}\\
\dot{I}&=\beta(t,I) S I-\gamma I, \label{eq:I_cmodel}\\
\dot{R}&=\gamma I, \label{eq:R_cmodel}
\end{align}
Here, $I$ is the overall infection number (all strains together) and
$\beta(t,I)$ is the average infection rate, which can depend on the overall infection number, depending on the underlying mutation statistics (see Methods).  

\begin{figure*}[!t]
    \centering
    \includegraphics[width=1.0\textwidth]{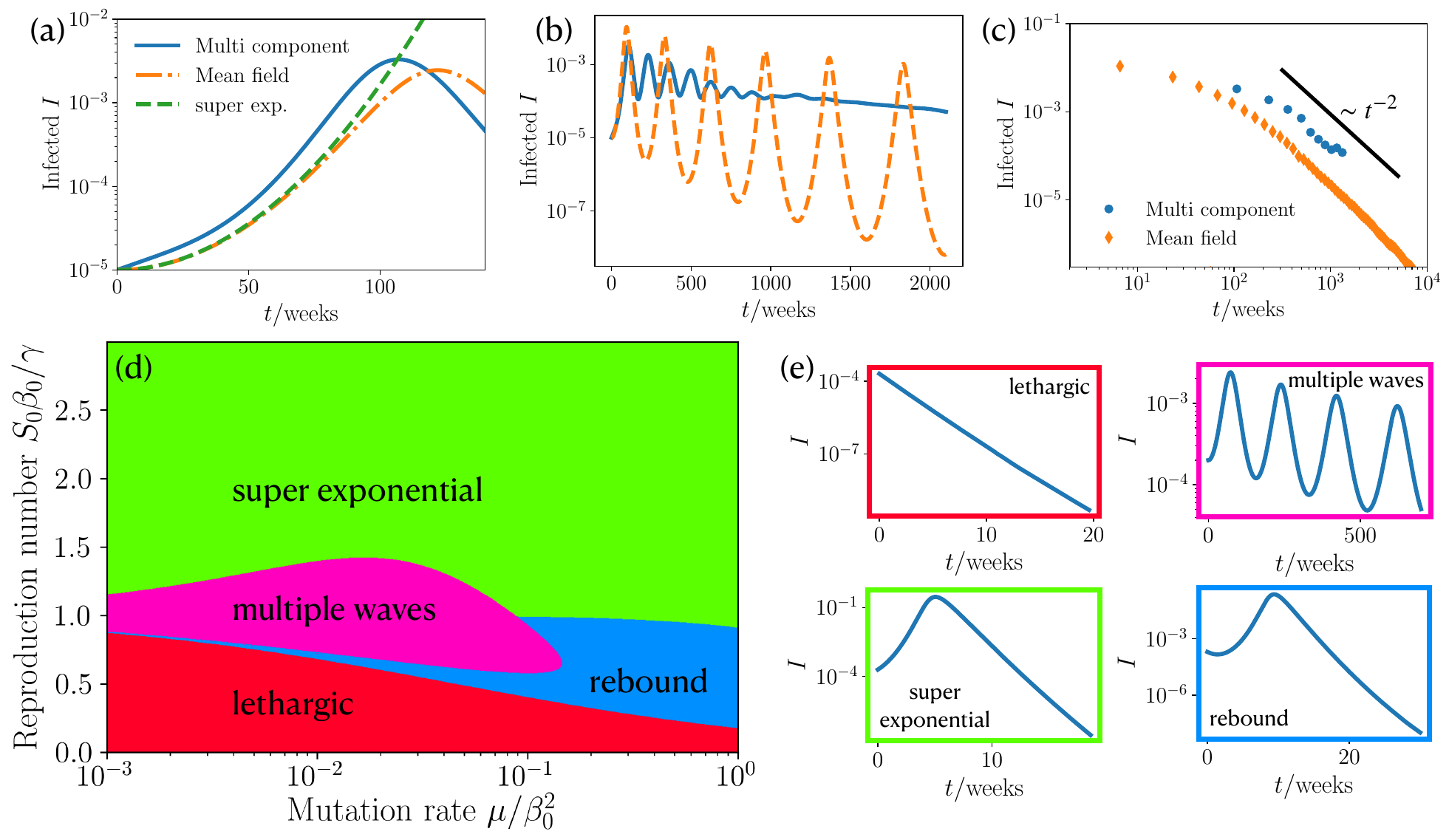}
    \caption{
    Power law dependence of infection dynamics, phase diagram and state classification for constant mutation rate. (a) Fraction of infected people $I$ at short times $t$ for the coarse grained MSIR, multi component MSIR and the early time approximation Eq.~\eqref{eq:I_early_times_i}. ($I_0=10^{-5}$, $R_0=1$, $\mu/\beta_0^2=0.05$) (b) Long time wave pattern of fraction of infected people for coarse grained MSIR and multi component MSIR approach. ($I_0=10^{-5}$, $R_0=1$, $\mu/\beta_0^2=0.2$) (c) Scaling of maxima of infections for coarse grained MSIR and multi component MSIR approach. ($I_0=10^{-5}$, $R_0=1$, $\mu/\beta_0^2=0.2$)
    (d) Phase diagram for our coarse grained infection dynamics showing the occurrence of four different courses of the pandemic for varying mutation rate and reproduction number. ($I_0=2\times10^{-4}$) (e) Different courses of infections during an epidemic: lethargic, multiple waves, super exponential, and rebound.}
    \label{fig:combinedMI}
\end{figure*}

\vskip 0.6cm 
\noindent\large \textbf{Results} \\ \normalsize
Let us now explore the impact of mutations on the disease evolution by comparing numerical simulations of the multi-component model with analytical predictions based on the mean-field model (Fig.~\ref{fig:model_sketch} (b)). 
To allow distinguishing between direct effects of mutations from the prevailing (and most infectious) strain and indirect effects due to the self-accelerating mutation cascade which we have described in the introduction and which may or may not become effective in reality, depending on the actual mutation rate and other parameters, we 
will sequentially follow on the cases of    
(i) a constant mutation rate $\mu$ leading to the emergence of new strains in our simulations with a constant rate 
and (ii) a mutation rate which depends on the present infection number $\mu(I)$.
%where the 
%mutation rate is not constant and coupled to the infection number $\nu$, as discussed in %the introduction. 
\vskip 0.5cm 
\noindent\textbf{Constant mutation rate}
\vskip 0.2cm \noindent
\emph{Mutation-driven infection dynamics --}
Let us assume that the infection rate $\beta_n$ of a newly occurring virus-strain 
is randomly selected from 
a normal distribution $p$ with standard deviation $\sigma$ centered around the infection rate of the presently prevailing strain: 
\begin{align}
    p(\beta_n)= \frac{1}{\sqrt{2\pi}\sigma} \mathrm{Exp}(-\frac{(\beta_n-\beta_{\mathrm{max},n-1})^2}{2\sigma^2}),
    \label{eq:Gaussian_dist_infect}
\end{align}
Here $\beta_{\mathrm{max},n-1}$ denotes the largest infection rate of all currently existing strains. This distribution, the average of which moves towards higher values in the course of a disease, is motivated 
by the fact that newly occurring mutations typically become visible only if they have a higher (or at least not much lower) infection rate than the currently prevailing ones. 
Coarse graining this mutation statistic (see Methods: Details on model beyond constant mutation rate) yields an the following average infection rate for our mean-field model %evolves according to the equation 
\begin{align}
    \beta= \beta_0 + \mu t
    \label{eq:constant_mutation_rate}
\end{align}
where $\mu$ is the constant mutation rate and $\beta_0$ is the initial infection rate.
That is, coarse graining the distribution (\ref{eq:Gaussian_dist_infect}) leads to a constant 
increase of the infection rate with time. 
%Here the mutations lead to a constant increase of the infection rate with time, meaning that %the disease becomes more infectious in time. 
\\Let us first explore the disease evolution at early times when the majority of the population is susceptible, such that $S(t) \approx 1$. Then Eq.~\eqref{eq:I_cmodel} reduces to
\begin{align}
    \dot{I}&=\beta(I,t) I-\gamma I. \label{eq:I_early_times}
\end{align}
Now using Eq.~\eqref{eq:constant_mutation_rate} we find
\begin{align}
    I(t)= I_0 \mathrm{exp}\left[ 
    \frac{1}{2} t (2\beta_0 -2 \gamma +t \mu )
    \right].
    \label{eq:I_early_times_i}
\end{align}
Thus, the fraction of infected individuals does not grow exponentially as in the standard SIR model but even faster. Following Eq.~\eqref{eq:I_early_times_i} if the mutation rate is high enough that $t\mu \gg 2(\beta_0-\gamma)$ long before herd-immunity is reached, the infection dynamics generically converges towards $I(t)\propto e^{\mu t^2}$, which is completely mutation-driven. 
To test this prediction, we now numerically solve the full multi-component model and show the overall $I(t)=\sum_n I_n(t)$ in 
Fig.~\ref{fig:combinedMI}(a). Notably, the result is close to the analytical prediction of the mean-field model and shows an even slightly larger growth. 

%A super exponential growth of infections in an epidemic is a particularly dangerous scenario, since this will also give a strong rise in hospitalized and diseased people.
\begin{figure*}[!t]
    \centering
    \includegraphics[width=1.0\textwidth]{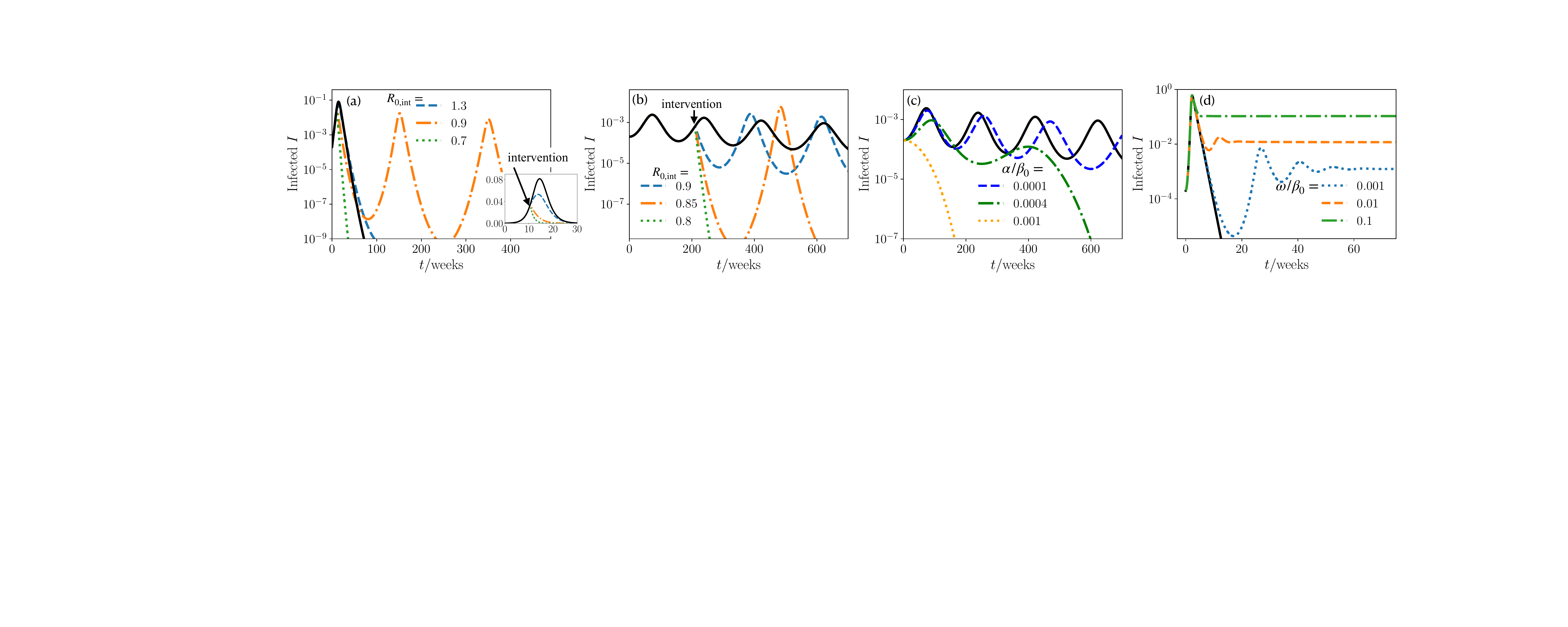}
    \caption{Nonpharmaceutical interventions, vaccinating, and  immune escape. (a) Infections as function of time for a super exponential wave. Measures are taken by reducing the reproduction number to $R_{0,\mathrm{int}}$ at time $t_{\mathrm{int}}$. Inset: zoom in to the early time regime. (Black line has no interventions, $R_0=1.5$, $\mu/\beta_0^2=0.004$, $t_{\mathrm{int}}=10.5$ weeks) (b) Infections as function of time for a wave like pandemic course. Measures are taken by reducing the reproduction number to $R_{0,\mathrm{int}}$ at time $t_{\mathrm{int}}$. (Black line has no interventions, $R_0=1$, $\mu/\beta_0^2=10^{-3}$, $t_{\mathrm{int}}=210$ weeks)
    (c) Infections as function of time of different vaccination rates $\alpha$. (Black line has no vaccination, $R_0=1$, $\mu/\beta_0^2=10^{-3}$, $I_0=2 \times 10^{-4}$)
    (d) Infections as function of time of different immune escape rates $\omega$. (Black line has no immune escape, $R_0=1.2$, $\mu/\beta_0^2=2$, $I_0=2 \times 10^{-4}$)}
    \label{fig:Model1_measure}
\end{figure*}
Clearly, the predicted (super)exponential growth of the infection numbers can not continue forever but has to saturate once the population reaches herd-immunity either by collectively going through the infection or through vaccination. Once herd-immunity is reached, the infection numbers are normally expected to monotonously decrease, as predicted by the standard SIR model. However, numerical solutions of our mean-field model show that after a phase where the population recovers and infection numbers decay to a very low level, they can rapidly grow again (Fig.~\ref{fig:combinedMI}(b)). This sequence of decreasing and increasing infection numbers can even repeat for many times, leading to an infection wave-train. The maxima of the wave-trains follow a scaling law of $I_{\mathrm{max}}\sim 1/(\beta_0+\mu t_{\mathrm{max}})^2$ (see SI for derivation), which is shown in Fig.~\ref{fig:combinedMI}(c).
The prediction of wave trains is also confirmed by numerical solutions of the multi-component model, but somewhat weakened, because the individual strains can show waves occurring at individual "frequencies". 
Let us now ask about the mechanism leading to these infection waves. 
They are induced by the nonlinear coupling of the infected and susceptible. First, the number of infected people grows, when the term $\beta S$ in Eq.~\eqref{eq:I_cmodel} is large enough. At a certain point, the number of susceptible is too small and the saturation effect from the recovery rate $\gamma$ in Eq.~\eqref{eq:I_cmodel} takes over such that the number of infected decreases. However, the infection rate $\beta$ continues growing with time, such that $\beta S$ can become large enough to induce a second wave. This feedback continues on multiple times giving rise to the oscillatory behavior shown in Fig.~\ref{fig:combinedMI}(b). 
One a more intuitive level, these considerations show that in the presence of mutations 
herd-immunity is not necessarily a persistent state of a population and that strongly decreasing infection numbers are not an overall reliable sign that the population has overcome the disease. From a sociopolitical viewpoint each growth phase within such wave trains might evoke (nonpharmaceutical) interventions, creating an immense mental burden on the population. 
\vskip 0.5cm 
\emph{Phase diagram --}
To see how strong measures have to be taken to 
prevent such an infection wave train (or a super-exponential growth) in the first place,  
we now systematically vary the parameters in the model to create a state diagram providing a systematic overview on the possible scenarios. 
It turns out that there are three dimensionless control parameters in our system (see SI), one of which is the initial infection number and the other two ones are 
the effective reproduction number $S_0\beta_0/\gamma$ and a dimensionless mutation rate $\mu/\beta_0^2$. Varying both parameters systematically and solving the mean-field mutation susceptible-infected-recovered (MSIR) model for each parameter combination we obtain the phase diagram shown in Fig. ~\ref{fig:combinedMI}(e), which shows four qualitatively different epidemic courses:
a lethargic phase, which is characterized by an exponential decay, multiple waves, super exponential wave, and a rebound, with an initial local minimum and a proceeding super exponential increase.
The occurrence of these states in parameter space is summarized in Fig.~\ref{fig:combinedMI}(d). At large reproduction number the dynamics is "super exponential" (green domain) for any positive mutation rate. When decreasing the reproduction number, depending on the mutation rate, one reaches the regime of "multiple waves" which we have previously discussed (pink) or a "rebound" phase (blue) where infection numbers initially decrease, pass a minimum and then increase to reach a single maximum before finally decreasing (Fig.~\ref{fig:combinedMI}(e)). For even lower mutation rates, the population is in the "lethargic" regime, where the infection numbers monotonously decrease.  
\vskip 0.5cm 
\emph{Nonpharmaceutical interventions, vaccination, and immune escape --}
In practice the goal is of course be to apply appropriate measures to safely reach the lethargic regime in Fig.~\ref{fig:combinedMI}(d) and not to end up in the multiple wave or rebound regime where the 
evolution of infection numbers show a promising initial trend but a severe evolution at later times.
To understand the impact of nonpharmaceutical interventions, we reduce the 
reproduction number~\cite{maier2020effective,dehning2020inferring,te2020effects,te2021containing,tian2021harnessing,meidan2021alternating},
from $R_0$ to a reduced reproduction number including measures $R_{0,\mathrm{int}}$ at a time $t_{\mathrm{int}}$ in our simulations and numerical solutions of the MSIR model.  
Starting in the super exponential wave regime ($R_0=1.5$) we apply interventions during the rise of a first wave (see Fig.~\ref{fig:Model1_measure}(a)).
By including weak measures ($R_{0,\mathrm{int}}=1.3$), the maximum of infections decreases as intended, however, the wave needs a longer time to decay, implying a longer period of restrictions for the public. Lowering the reproduction number to $R_{0,\mathrm{int}}=0.9$, results in the appearance of a second and third intervention-mutation induced wave. These waves are enabled by the increased infection rate and the fact that due to the interventions there are more susceptible at a later point in time, where they can facilitate the growth of infections.
Here, the situation would be particularly confounding to the public, since it was subjected to measures to decrease the number of infections in the first place; however, this results in more waves and a likely extension of the period of interventions. On the other hand, a strong reduction of the reproduction number ($R_{0,\mathrm{int}}=0.7$) gives a fast decay of infections, as intended. 
Of course the situation changes when we start from an infection dynamics with multiple waves ($R_0=1$) and apply measures during the rise of the second wave (see Fig.~\ref{fig:Model1_measure}(b)). As expected, strong measures ($R_{0,\mathrm{int}}=0.8$) have the intended effect of eliminating the epidemic. On the other hand, if the measures are slightly weaker ($R_{0,\mathrm{int}}=0.85$), the infections first decrease, but then lead to a second intervention-mutation induced delayed wave, which is stronger in magnitude than the first wave. Again, this wave is induced by the growing infection rate and the enhanced number of susceptible individuals due to interventions. To decision makers and the public this type of wave could likely appear as unexpected. 
However, note that at least the overall number of recovered people at the end of our the epidemic is decreasing with stronger interventions, meaning that if only the cumulative number of infections is considered every reduction of $R_0$ is useful.

\begin{figure*}[!t]
    \centering
    \includegraphics[width=1.0\textwidth]{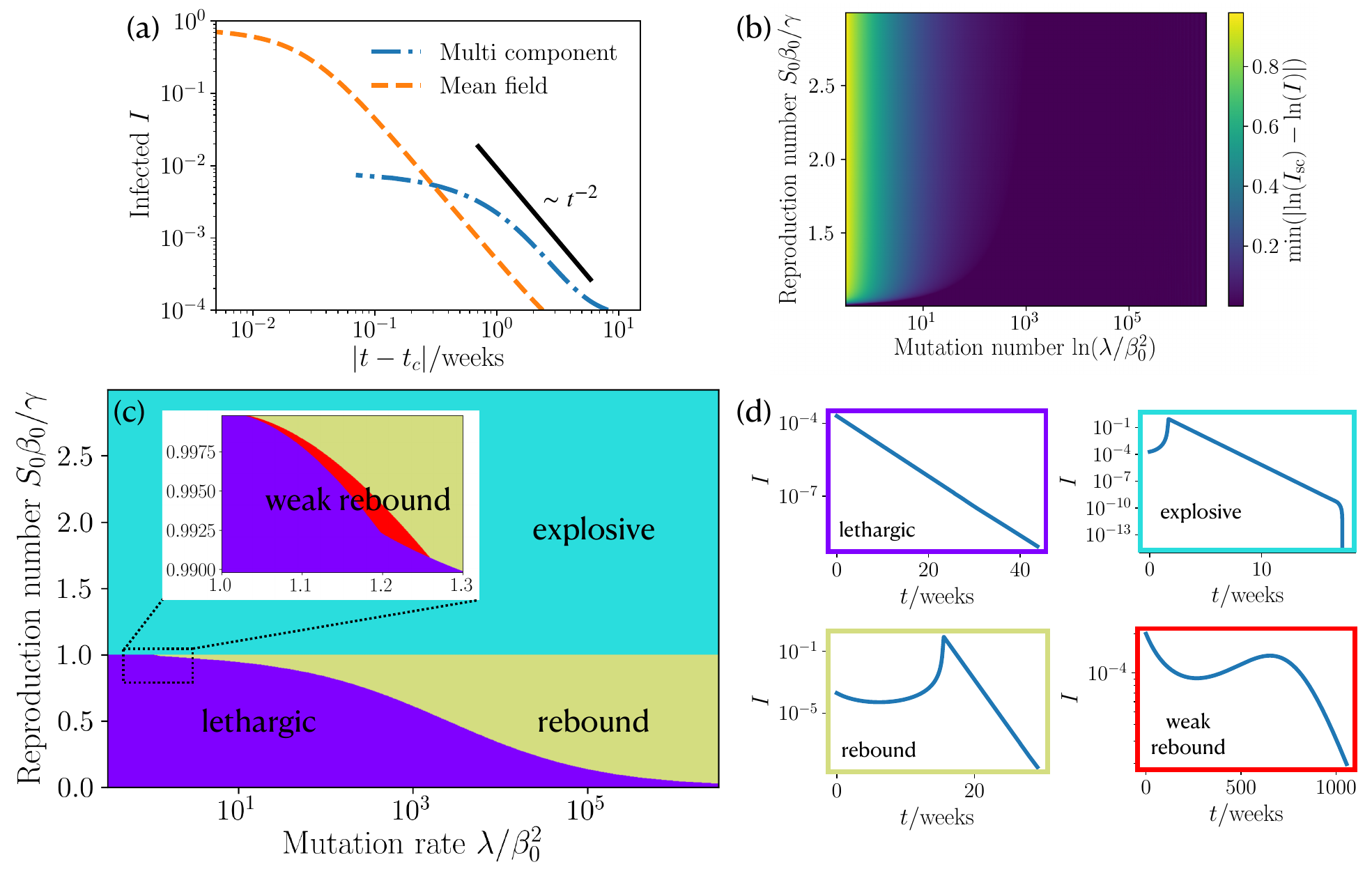}
    \caption{Scaling law of short time infection dynamics, Phase diagram and state classification of approach beyond constant mutation rate. (a) Infections as function of reduced time $|t-t_c|$ where $t_c$ is the critical time at which the infections diverge (see SI for details). We show the scaling law, coarse grained MSIR and our multi component MSIR approach. ($R_0=2$,  $\lambda/\beta_0^2 = 2\times 10^4$, $I_0=10^{-4}$) (b) Deviation of the fraction of infections at short times from our corse grained MSIR approach to the a $1/|t-t_c|^2$ scaling by using $\mathrm{min}(\mathrm{ln}(I_{\mathrm{sc}}) - \mathrm{ln}(I))$. Mutation rate and reproduction number are varied.  ($I_0= 2\times 10^{-4}$)
   (c) Phase diagram of our coarse grained MSIR approach showing the occurrence of four different courses of the pandemic for varying mutation rate and reproduction number. ($I_0= 2\times 10^{-4}$) (d) Example plots of the infections as function of time for four different courses of the pandemic: lethargic, super exponential wave; rebound, and weak rebound.}
    \label{fig:combinedMII}
\end{figure*}
Specifically for COVID-19 vaccines have become available and their continuous production gives hope to get the disease under control. % overcome the disease soon. 
However, one and half a years after vaccines have first become available only about 67\% of the worldwide population has been fully vaccinated (early Mai 2022), leaving much time for the emergence of highly infectious mutations. 
Vaccinations effectively reduce the number of  susceptible in the SIR model~\cite{grauer2020strategic}, such that we modify Eq.~\eqref{eq:S_cmodel} as
\begin{align}
    \dot{S}&=-\beta(I,t) S I - \alpha, \label{eq:S_cmodel_vaccine}
\end{align}
with a vaccination rate $\alpha$. Note that a more realistic model for vaccination would also account for a time dependent roll out.
We investigate the effect of vaccination on the infection dynamics for a situation leading to multiple waves (see Fig.~\ref{fig:Model1_measure}(c)). 
Following Fig. 3(c) vaccinations have a clear effect; they have to be applied fast enough to significantly reduce the number of infections and temper the train of waves. This further shows the importance of manufacturing and distributing vaccinations as fast as possible. 

When newly mutated strains arise the possibility of an immune escape increases, where a recovered individual does not stay immune against the new strain. Let us therefore now include a term which accounts for an effective immune escape in our model in the form of a transition from recovered to susceptible. Explicitly we modify Eqs.~\eqref{eq:S_cmodel}-\eqref{eq:R_cmodel} to read
\begin{align}
\dot{S}&=-\beta(t,I) S I + \omega R, \label{eq:S_cmodel_immune}\\
\dot{I}&=\beta(t,I) S I-\gamma I, \label{eq:I_cmodel_immune}\\
\dot{R}&=\gamma I - \omega R, \label{eq:R_cmodel_immune}
\end{align}
where $\omega$ is the immune escape rate. The transition rate between susceptible and recovered introduces two new effects. 
First, this term leads to a positive feedback loop between the number of susceptible and recovered individuals which induces infection waves (see Fig.~\ref{fig:Model1_measure}(d)), which can also be seen from a linear stability analysis (see SI) of Eqs.~\eqref{eq:S_cmodel_immune}-\eqref{eq:R_cmodel_immune}. Second, the infected population fraction saturates to a nonzero steady state, due to a replenishment of susceptibles (see also SI).

\vskip 1cm 
\noindent\textbf{Beyond constant mutation rate} \\
%Let us now account for the fact that in reality the mutation rate is coupled to the infection number,
As a second possible scenario, let us now assume that the mutation rate is coupled to the infection number,
such that mutations are more likely in phases where the infection numbers are large. To account for this effect in our model, we 
assume that new (relevant) mutations occur with a probability of $p_0 I_{n-1}$ from the most infectious strain where $p_0$ is constant and also that the infection rates of new strains follow from a random walk with a mutation-induced bias as 
$\beta_n= \beta_{n-1} + \Delta \beta$ (see Methods for details). 
%are likely to be more infectious than proceeding strains. 
Coarse graining the biased random walk yields a mean field infection rate which evolves as
\begin{align}
    \beta(t)= \int_0^t \lambda I(t') \mathrm{d} t',
    \label{eq:coarse_grain_rate_random_walk}
\end{align}
with a mutation rate $\lambda$. Intuitively, this means that new mutations occur in our mean-field model with a rate which is proportional to the present infection number. 
\vskip 0.5cm 
\emph{Mutation-induced dynamics --}
 At early times, where $S\approx 1$ we obtain again Eq.~\eqref{eq:I_early_times}, which yields together with Eq.~\eqref{eq:coarse_grain_rate_random_walk}
\begin{align}
    I(t)=-\frac{\delta_1}{\lambda\left[1+\cosh \left(2 \ln \delta_2 +t \sqrt{\delta_1}\right)\right]},
    \label{eq:I_early_times_ii}
\end{align}
where $\delta_1$ and $\delta_2$ are constants that are given in the SI. Importantly, the infections in Eq.~\eqref{eq:I_early_times_ii}  do not grow exponentially, but there is an explosive super exponential growth, which asymptotically has a scaling behavior following $I_{\mathrm{sc}}(t) \sim 1/|t-t_c|^2$ with a critical time $t_c$ (see SI for explicit expression). Crucially, this giant infection wave qualitatively differs from the comparatively mild super-exponential behaviour which we have encountered for the case of a constant mutation rate in that it leads to a much more extreme self-acceleration of the infection numbers. As a result, the infection numbers peak only at extremely high values where a large fraction of the population is infected at the same time (see Fig. 1b and 
Fig.~\ref{fig:combinedMII}(a)). 
Clearly, such an explosive growth would be interrupted at some point 
as the population approaches herd immunity. To quantify to which extend the predicted explosive growth would occur before herd-immunity causes significant deviations, we consider 
the expression $\mathrm{min}(\mathrm{ln}(I_{\mathrm{sc}}) - \mathrm{ln}(I))$, which quantifies how closely the fraction of infections approaches the underlying (idealized) power law dependence in the presence of saturation effects. 
We find that for large mutation rates the power-law dependence is strong for any reproduction number (Fig.~\ref{fig:combinedMII}(b)) and weakens for lower mutation rates. 
Remarkably, the explosive growth depends only weakly on the reproduction 
which is the parameter that is controllable due to interventions.
After the initial super-exponential increase of infections the saturation effects from people recovering induce a maximum and a succeeding decrease in infection numbers. 
\vskip 0.5cm 
\emph{Phase diagram --} Depending on the basic reproduction number and the mutation rate the MSIR model predicts four distinct courses summarized in the phase diagram Fig.~\ref{fig:combinedMII}(c), which has been obtained analytically (see SI). We find a lethargic regime characterized by an exponential decay (purple regime, Fig.~\ref{fig:combinedMII}(d)); an explosive regime (cyan regime); a rebound regime (dark yellow) where we have a minimum followed by a mutant induced super exponential increase; and weak rebound (red) leading to an infection maximum which is smaller than the initial fraction of infected individuals. 
Generally, the explosive (or super exponential) regime occurs for 
reproduction numbers $R_0>1$ and any positive mutation rate, whereas the other three regimes occur for 
$R_0<1$. For low mutation rate and $R_0<1$ the epidemic is in the desired lethargic regime, increasing the mutation rate leads to a small region of weak rebound which then transitions to a rebound dynamics. 
%This state diagram might in principle help navigating through an epidemic, showing how lowering the reproduction number induces a different state.

\begin{figure*}[!t]
    \centering
    \includegraphics[width=1.0\textwidth]{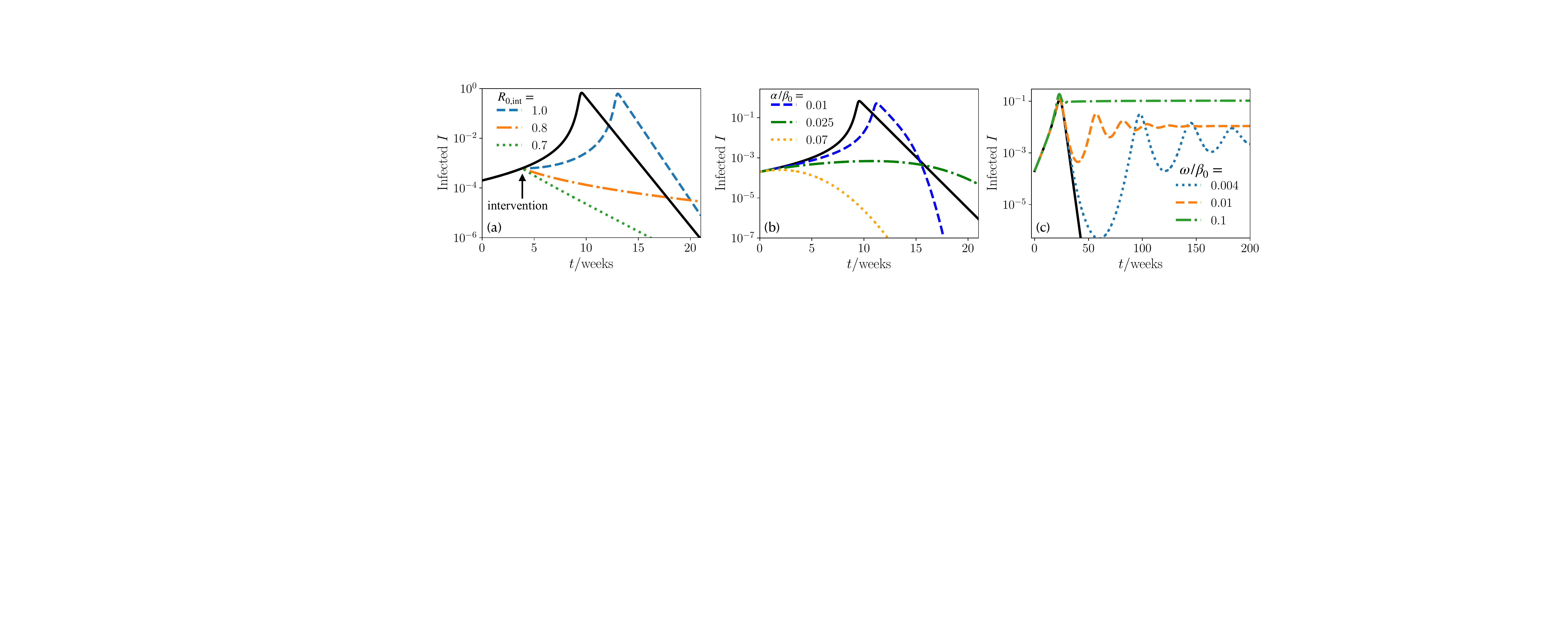}
    \caption{Nonpharmaceutical interventions, vaccinating, and  immune escape (a) Infections as function of time for a super exponential wave. Measures are taken by reducing the reproduction number to $R_{0,\mathrm{int}}$ at time $t_{\mathrm{int}}$. (Black line  has no intervention. $R_0=1.2$, $\lambda/\beta_0^2=50$,  $I_0=2\times10^{-4}$,  $t_{\mathrm{int}}=3.85$ weeks)
    (b) Infections as function of time of different vaccination rates $\alpha$. (Black line has no vaccination. $R_0=1.2$, $\lambda/\beta_0^2=50$, $I_0=2\times10^{-4}$)
    (c) Infections as function of time of different immune escape rates $\omega$. (Black line has no immune escape. $R_0=1.2$, $\lambda/\beta_0^2=2$, $I_0=2\times10^{-4}$)}
    \label{fig:Model2_measure}
\end{figure*}

\vskip 0.5cm 
\emph{Nonpharmaceutical interventions, vaccination, and immune escape --}
To explore the efficiency of measures which effectively reduce the reproduction number, we again  
change the reproduction number $R_0$ to a value of $R_{0,\mathrm{int}}$ at time $t_{\mathrm{int}}$. 
Now starting from the explosive (super exponential) regime ($R_0=1.2$) we decrease the reproduction number during the rise of the wave (Fig.~\ref{fig:Model2_measure}). Strong measures yielding $R_{0,\mathrm{int}}=0.7$ or $R_{0,\mathrm{int}}=0.8$) induce an immediate decay of the infection numbers as desired. However, weak measures, leading to $R_{0,\mathrm{int}}=1$ only delay the occurrence of the infection maximum but hardly change the value of $I$ at the peak. Hence, it is clear that measures need to be strong enough to have a significant effect, while weak measures only delay the infection number explosion. 

%In view of the ongoing vaccination campaign against COVID-19 
We finally ask how fast vaccination would have to progress in order to suppress a mutation-induced explosion of infection numbers or a mutation-induced rebound at initial reproduction numbers smaller than 1. 
Vaccinations effectively reduce the number of 
susceptibles by a vaccination rate $\alpha$ (see also Eq.~\ref{eq:S_cmodel_vaccine}).
If we start in the explosive (super exponential) regime, a high vaccination rate is needed to interrupt the rapid growth of the infection numbers (Fig.~\ref{fig:Model2_measure}(b)): while 
small vaccination rates ($\alpha/\beta_0=0.01$) merely shift the infection maximum to a later time, while having little change in the number of infections, larger vaccination rates ($\alpha/\beta_0=0.025$) 
effectively suppress the explosion of infection numbers. 
Therefore, to prevent possible mutation-induced long-time consequences it is imperative to maximize vaccine production worldwide. redEnhancing vaccine production is particularly important in countries with a weak healthcare system, which face the threat of being overloaded by COVID-19 cases \cite{hasan2021emergence,essar2021covid,islam2022food,aborode2021impact,aborode2021dengue,rackimuthu2021covid}.

Newly mutated strains allow for an immune escape of the virus, effectively representing a transition of recovered individuals to susceptible ones with an immune escape rate $\omega$ (see also Eqs.~\eqref{eq:S_cmodel_immune}-\eqref{eq:R_cmodel_immune}). This leads to a continued recovery of the number of susceptibles and drives the population away from herd immunity, which in turn can cause new infection waves (see Fig.~\ref{fig:Model2_measure}(c)), that can also be predicted using a linear stability analysis (see SI). Further, we find that the number of infected saturates to a nonzero steady state (see also SI).

\vspace{1cm}
\noindent\large \textbf{Discussion} \\ \normalsize
Inspired by the ongoing COVID-19 disease and the continued emergence of new mutations supplanting preceding ones, in the present work we have developed a stochastic multistrain generalization of the popular SIR model. Combining this model with coarse graining concepts from statistical physics has allowed us to predict a panorama of possible scenarios for the mutation-controlled evolution of infectious diseases. 

In particular, our approach suggests that mutations can induce a super-exponential growth of infection numbers in populations which are highly susceptible to the disease (e.g. because they are far from reaching herd immunity).
%which are still from reaching herd-immunity. 
As compared to the standard exponential growth, 
interrupting such an super-exponential growth is much more difficult and requires stronger and stronger measures as the disease evolves. In practice, such a super-exponential growth may occur e.g. 
if measures are applied too late, or if vaccines suddenly become ineffective against mutations. 

One particularly severe form of such an super-exponential growth can occur 
if the mutation rate of a virus is proportional to the current number of 
infections. For this case our model predicts a giant infection wave, which is based on a 
positive feedback loop between the mutation-rate and the infection number causing a massive-self acceleration of the latter resulting in a state where the 
majority of the population gets infected at the same time. Clearly, such a situation would not only massively overstress any existing health system but once in action it would hardly be interruptable through vaccination.  

At later stages of an infectious disease, where the population approaches herd immunity and the infection numbers decrease, an obvious political reaction would be to release measures. However, our simulations suggest that
mutations can drive new infection waves even after a longer downwards trend. Such waves can even self-repeat and 
lead to a pattern of repeated phases of strongly decaying and increasing infection numbers provoking an endless sequence of renewed non-pharmaceutical interventions.

Since our work is on a conceptual basis we did not include explicit data to model COVID-19. However, the panorama of mutation induced phenomena which we have identified mightf inspire detailed modeling 
works to test them for specific infectious diseases such as COVID-19. red Further, our results could be applied to diseases in the animal world such as the avian influenza~\cite{o2013chaos}.
%While our work is on a conceptual basis, this panorama of mutation induced phenomena which we have identified should inspire detailed modeling works to test them for specific infectious diseases such as COVID-19. 
These results might also be useful for discussions regarding the importance of a release of vaccine-patents to reduce the risk of mutation-induced infection revivals and to coordinate the release of measures following a downwards trend of infection numbers.

\vspace{1cm}
\noindent \textbf{Methods}
\\
\small
\textbf{Basic reproduction number of COVID-19 mutants.}
The basic reproduction numbers where extracted from: original  variant~\cite{liu2020reproductive}, B.1.1.7~($\alpha$)~\cite{volz2021assessing}, B.1.351~($\beta$)~\cite{assessment2021risk},
P.1~($\gamma$)~\cite{coutinho2021model}, 	B.1.617.2~($\delta$)~\cite{campbell2021increased}, and B.1.1.529~($o$)~\cite{nishiura2021relative}. The time point at which a variant has reached $5\%$ in the sequenced genomes reported  in~\cite{hadfield2018nextstrain} (\url{https://nextstrain.org/ncov/global}) is used as the emergence time. 
\\
\textbf{Details on constant mutation rate}
We assume that the new infection rates $\beta_n$ are drawn from a Gaussian distribution, whose mean is the largest current infection rate. Explicitly we have
\begin{align}
    p(\beta_n)= \frac{1}{\sqrt{2\pi}\sigma} \mathrm{Exp}(-\frac{(\beta_n-\beta_{\mathrm{max},n-1})^2}{2\sigma^2}),
    \label{eq:Methods_Gaussian_dist_infect}
\end{align}
where $\beta_{\mathrm{max},n-1}$ denotes the maximal infection rate of the current strains, $\beta_n$ is the infection rate of the newly mutated strain, $\sigma$ is the standard deviation of the distribution, and new strains are  produced at a rate $m$. (In our multi component simulations we use $\beta_0/\gamma=1$, $\sigma=2\times10^{-4}$, $m/\gamma=  2$, and the new strain obtains an initial $I_n(0)= 10^{-7}$)

To coarse grain this mutation model, we assume that the infections immediately assume the maximal infection rate of the newly mutated strain $\beta_{\mathrm{max},n}$. It follows that the mean infection rate is dominated by $\beta_{\mathrm{max},n}$, since all other infections grow exponentially slower. This reduces our multicomponent model to an effective one component model with the infection rate $\beta_{\mathrm{max},n}$. To determine $\beta_{\mathrm{max},n}$ we compute
\begin{align}
    \int_{\mathrm{max}_n\beta_{n}}^{\infty} p(\beta_n) \mathrm{d}\beta_n = \frac{\gamma}{m},
    \label{eq:beta_max_cond}
\end{align}
where $m$ is the number of times drawn from the distribution Eq.~\eqref{eq:Methods_Gaussian_dist_infect}.
Explicitly, Eq.~\eqref{eq:beta_max_cond} yields
\begin{align}
    \beta_{\mathrm{max},n}= \beta_{\mathrm{max},n-1} -\sqrt{2} \sigma \mathrm{erf}^{-1}(-1+\frac{2\gamma}{m}),
    \label{eq:beta_max_sol}
\end{align}
where $\mathrm{erf}^{-1}(*)$ is the inverse error function and can here be approximated by a negative constant $-C_1$. We now write the standard deviation as $\sigma= \mu^* \tau$, with a mutation rate $\mu^*$ and mutation timescale $\tau$, giving
\begin{align}
        \frac{\beta_{\mathrm{max},n}- \beta_{\mathrm{max},n-1}}{\tau}= C_1 \sqrt{2} \mu^*,
    \label{eq:Methods_beta_diff}
\end{align}
which is a discretized version of $\dot \beta= \mu$, and equivalent to our coarse grained  constant mutation rate model.

\noindent
\textbf{Details on model beyond constant mutation rate}
For model beyond constant mutation rate the infection rates perform a biased random walk. Given an infection rate $\beta_{n}$ it will mutate with a probability $p_0$ and not mutate with probability $1-p_0$. Furthermore, this strain has $I_n$ infections, which are all able to mutate, giving a total mutation probability of $1-(1-p_0)^{I_n}= p_0 I_n +\mathcal{O}(p_0^2)$. A mutation gives a new strain $I_n$ with an increased $\beta_n= \beta_{n-1}+ \Delta \beta$. (In our multi component simulations we use $\beta_0=0.1$, $\gamma=0.1$, $\Delta \beta=0.03$, $p_0=2\times 10^{-4}$, and the new strain obtains an initial $I_n(0)= 10^{-6}$)

To coarse grain, we assume that the expectation value of the infection rate $\langle \beta_n \rangle$ is proportional to the mutation probability. Then a new mutation has the expectation value $\langle \beta_{n+1} \rangle= p_0 I_n$ and the old strain has $\langle \beta_{n} \rangle= 1- p_0 I_n$. Computing the difference gives
\begin{align}
    \langle \beta_{n+1} \rangle- \langle \beta_{n} \rangle= 2 p_0 I_n -1,
\end{align}
which is a discretization of $\dot \beta = \lambda I$, which is our coarse grained model beyond a constant mutation rate.

\normalsize
\vspace{1cm}
\noindent \textbf{Author contributions}
\small
H.L. and B.L. designed the research. J.G. and F.J.S. performed the simulations. F.J.S. did the analytical calculations and the simulation data analysis. 
All authors discussed and wrote the manuscript.

\vspace{1cm}\normalsize
\noindent\textbf{Competing interests}
\small
The authors declare no competing interests.

%\bibliography{references}% Produces the bibliography via BibTeX.

\begin{thebibliography}{10}
\urlstyle{rm}
\expandafter\ifx\csname url\endcsname\relax
  \def\url#1{\texttt{#1}}\fi
\expandafter\ifx\csname urlprefix\endcsname\relax\def\urlprefix{URL }\fi
\expandafter\ifx\csname doiprefix\endcsname\relax\def\doiprefix{DOI: }\fi
\providecommand{\bibinfo}[2]{#2}
\providecommand{\eprint}[2][]{\url{#2}}

\bibitem{zhou2020pneumonia}
\bibinfo{author}{Zhou, P.} \emph{et~al.}
\newblock \bibinfo{journal}{\bibinfo{title}{A pneumonia outbreak associated
  with a new coronavirus of probable bat origin}}.
\newblock {\emph{\JournalTitle{Nature}}} \textbf{\bibinfo{volume}{579}},
  \bibinfo{pages}{270--273} (\bibinfo{year}{2020}).

\bibitem{wu2020new}
\bibinfo{author}{Wu, F.} \emph{et~al.}
\newblock \bibinfo{journal}{\bibinfo{title}{A new coronavirus associated with
  human respiratory disease in china}}.
\newblock {\emph{\JournalTitle{Nature}}} \textbf{\bibinfo{volume}{579}},
  \bibinfo{pages}{265--269} (\bibinfo{year}{2020}).

\bibitem{Dong2020}
\bibinfo{author}{Dong~E, G.~L., Du~H}.
\newblock \bibinfo{journal}{\bibinfo{title}{An interactive web-based dashboard
  to track covid-19 in real time. (retrieved data on may 5th 2022 at
  \url{https://coronavirus.jhu.edu/})}}.
\newblock {\emph{\JournalTitle{Lancet Inf Dis.}}}
  \textbf{\bibinfo{volume}{20}}, \bibinfo{pages}{533--534},
  \doiprefix\url{10.1016/S1473-3099(20)30120-1} (\bibinfo{year}{2020}).

\bibitem{hadfield2018nextstrain}
\bibinfo{author}{Hadfield, J.} \emph{et~al.}
\newblock \bibinfo{journal}{\bibinfo{title}{Nextstrain: real-time tracking of
  pathogen evolution}}.
\newblock {\emph{\JournalTitle{Bioinformatics}}} \textbf{\bibinfo{volume}{34}},
  \bibinfo{pages}{4121--4123} (\bibinfo{year}{2018}).

\bibitem{rambaut2020dynamic}
\bibinfo{author}{Rambaut, A.} \emph{et~al.}
\newblock \bibinfo{journal}{\bibinfo{title}{A dynamic nomenclature proposal for
  sars-cov-2 lineages to assist genomic epidemiology}}.
\newblock {\emph{\JournalTitle{Nature Microbiology}}}
  \textbf{\bibinfo{volume}{5}}, \bibinfo{pages}{1403--1407}
  (\bibinfo{year}{2020}).

\bibitem{yao2020patient}
\bibinfo{author}{Yao, H.} \emph{et~al.}
\newblock \bibinfo{journal}{\bibinfo{title}{Patient-derived sars-cov-2
  mutations impact viral replication dynamics and infectivity in vitro and with
  clinical implications in vivo}}.
\newblock {\emph{\JournalTitle{Cell discovery}}} \textbf{\bibinfo{volume}{6}},
  \bibinfo{pages}{76} (\bibinfo{year}{2020}).

\bibitem{korber2020tracking}
\bibinfo{author}{Korber, B.} \emph{et~al.}
\newblock \bibinfo{journal}{\bibinfo{title}{Tracking changes in sars-cov-2
  spike: evidence that d614g increases infectivity of the covid-19 virus}}.
\newblock {\emph{\JournalTitle{Cell}}} \textbf{\bibinfo{volume}{182}},
  \bibinfo{pages}{812--827} (\bibinfo{year}{2020}).

\bibitem{grubaugh2020making}
\bibinfo{author}{Grubaugh, N.~D.}, \bibinfo{author}{Hanage, W.~P.} \&
  \bibinfo{author}{Rasmussen, A.~L.}
\newblock \bibinfo{journal}{\bibinfo{title}{Making sense of mutation: what
  d614g means for the covid-19 pandemic remains unclear}}.
\newblock {\emph{\JournalTitle{Cell}}} \textbf{\bibinfo{volume}{182}},
  \bibinfo{pages}{794--795} (\bibinfo{year}{2020}).

\bibitem{tegally2020emergence}
\bibinfo{author}{Tegally, H.} \emph{et~al.}
\newblock \bibinfo{journal}{\bibinfo{title}{Emergence and rapid spread of a new
  severe acute respiratory syndrome-related coronavirus 2 (sars-cov-2) lineage
  with multiple spike mutations in south africa}}.
\newblock {\emph{\JournalTitle{medRxiv}}}  (\bibinfo{year}{2020}).

\bibitem{priesemann2021action}
\bibinfo{author}{Priesemann, V.} \emph{et~al.}
\newblock \bibinfo{journal}{\bibinfo{title}{An action plan for pan-european
  defence against new sars-cov-2 variants}}.
\newblock {\emph{\JournalTitle{The Lancet}}} \textbf{\bibinfo{volume}{397}},
  \bibinfo{pages}{469--470} (\bibinfo{year}{2021}).

\bibitem{volz2021assessing}
\bibinfo{author}{Volz, E.} \emph{et~al.}
\newblock \bibinfo{journal}{\bibinfo{title}{Assessing transmissibility of
  sars-cov-2 lineage b. 1.1. 7 in england}}.
\newblock {\emph{\JournalTitle{Nature}}} \textbf{\bibinfo{volume}{593}},
  \bibinfo{pages}{266--–269} (\bibinfo{year}{2021}).

\bibitem{davies2021estimated}
\bibinfo{author}{Davies, N.~G.} \emph{et~al.}
\newblock \bibinfo{journal}{\bibinfo{title}{Estimated transmissibility and
  impact of sars-cov-2 lineage b. 1.1. 7 in england}}.
\newblock {\emph{\JournalTitle{Science}}} \textbf{\bibinfo{volume}{372}}
  (\bibinfo{year}{2021}).

\bibitem{coutinho2021model}
\bibinfo{author}{Coutinho, R.~M.} \emph{et~al.}
\newblock \bibinfo{journal}{\bibinfo{title}{Model-based evaluation of
  transmissibility and reinfection for the p. 1 variant of the sars-cov-2}}.
\newblock {\emph{\JournalTitle{medRxiv}}}  (\bibinfo{year}{2021}).

\bibitem{grauer2020strategic}
\bibinfo{author}{Grauer, J.}, \bibinfo{author}{L{\"o}wen, H.} \&
  \bibinfo{author}{Liebchen, B.}
\newblock \bibinfo{journal}{\bibinfo{title}{Strategic spatiotemporal vaccine
  distribution increases the survival rate in an infectious disease like
  covid-19}}.
\newblock {\emph{\JournalTitle{Scientific reports}}}
  \textbf{\bibinfo{volume}{10}}, \bibinfo{pages}{21594} (\bibinfo{year}{2020}).

\bibitem{zhouoptimizing}
\bibinfo{author}{Zhou, S.}, \bibinfo{author}{Zhou, S.}, \bibinfo{author}{Zheng,
  Z.} \& \bibinfo{author}{Lu, J.}
\newblock \bibinfo{journal}{\bibinfo{title}{Optimizing spatial allocation of
  covid-19 vaccine by agent-based spatiotemporal simulations}}.
\newblock {\emph{\JournalTitle{GeoHealth}}} \bibinfo{pages}{e2021GH000427}.

\bibitem{molla2021adaptive}
\bibinfo{author}{Molla, J.} \emph{et~al.}
\newblock \bibinfo{journal}{\bibinfo{title}{Adaptive and optimized covid-19
  vaccination strategies across geographical regions and age group}}.
\newblock {\emph{\JournalTitle{arXiv preprint arXiv:2105.11562}}}
  (\bibinfo{year}{2021}).

\bibitem{kermack1927contribution}
\bibinfo{author}{Kermack, W.~O.} \& \bibinfo{author}{McKendrick, A.~G.}
\newblock \bibinfo{journal}{\bibinfo{title}{A contribution to the mathematical
  theory of epidemics}}.
\newblock {\emph{\JournalTitle{Proceedings of the royal society of london.
  Series A, Containing papers of a mathematical and physical character}}}
  \textbf{\bibinfo{volume}{115}}, \bibinfo{pages}{700--721}
  (\bibinfo{year}{1927}).

\bibitem{hethcote2000mathematics}
\bibinfo{author}{Hethcote, H.~W.}
\newblock \bibinfo{journal}{\bibinfo{title}{The mathematics of infectious
  diseases}}.
\newblock {\emph{\JournalTitle{SIAM review}}} \textbf{\bibinfo{volume}{42}},
  \bibinfo{pages}{599--653} (\bibinfo{year}{2000}).

\bibitem{andersson2012stochastic}
\bibinfo{author}{Andersson, H.} \& \bibinfo{author}{Britton, T.}
\newblock \emph{\bibinfo{title}{Stochastic epidemic models and their
  statistical analysis}}, vol. \bibinfo{volume}{151}
  (\bibinfo{publisher}{Springer Science \& Business Media},
  \bibinfo{year}{2012}).

\bibitem{grassly2008mathematical}
\bibinfo{author}{Grassly, N.~C.} \& \bibinfo{author}{Fraser, C.}
\newblock \bibinfo{journal}{\bibinfo{title}{Mathematical models of infectious
  disease transmission}}.
\newblock {\emph{\JournalTitle{Nature Reviews Microbiology}}}
  \textbf{\bibinfo{volume}{6}}, \bibinfo{pages}{477--487}
  (\bibinfo{year}{2008}).

\bibitem{Gog17209}
\bibinfo{author}{Gog, J.~R.} \& \bibinfo{author}{Grenfell, B.~T.}
\newblock \bibinfo{journal}{\bibinfo{title}{Dynamics and selection of
  many-strain pathogens}}.
\newblock {\emph{\JournalTitle{Proc. Natl. Acad. Sci. USA}}}
  \textbf{\bibinfo{volume}{99}}, \bibinfo{pages}{17209--17214},
  \doiprefix\url{10.1073/pnas.252512799} (\bibinfo{year}{2002}).
\newblock \eprint{https://www.pnas.org/content/99/26/17209.full.pdf}.

\bibitem{harko2014exact}
\bibinfo{author}{Harko, T.}, \bibinfo{author}{Lobo, F.~S.} \&
  \bibinfo{author}{Mak, M.}
\newblock \bibinfo{journal}{\bibinfo{title}{Exact analytical solutions of the
  susceptible-infected-recovered (sir) epidemic model and of the sir model with
  equal death and birth rates}}.
\newblock {\emph{\JournalTitle{Applied Mathematics and Computation}}}
  \textbf{\bibinfo{volume}{236}}, \bibinfo{pages}{184--194}
  (\bibinfo{year}{2014}).

\bibitem{kroger2020analytical}
\bibinfo{author}{Kr{\"o}ger, M.} \& \bibinfo{author}{Schlickeiser, R.}
\newblock \bibinfo{journal}{\bibinfo{title}{Analytical solution of the
  sir-model for the temporal evolution of epidemics. part a: time-independent
  reproduction factor}}.
\newblock {\emph{\JournalTitle{J. Phys. A: Math. Theor.}}}
  \textbf{\bibinfo{volume}{53}}, \bibinfo{pages}{505601}
  (\bibinfo{year}{2020}).

\bibitem{schlickeiser2021analytical}
\bibinfo{author}{Schlickeiser, R.} \& \bibinfo{author}{Kr{\"o}ger, M.}
\newblock \bibinfo{journal}{\bibinfo{title}{Analytical solution of the
  sir-model for the temporal evolution of epidemics: part b. semi-time case}}.
\newblock {\emph{\JournalTitle{J. Phys. A: Math. Theor.}}}
  \textbf{\bibinfo{volume}{54}}, \bibinfo{pages}{175601}
  (\bibinfo{year}{2021}).

\bibitem{bittihn2020stochastic}
\bibinfo{author}{Bittihn, P.} \& \bibinfo{author}{Golestanian, R.}
\newblock \bibinfo{journal}{\bibinfo{title}{Stochastic effects on the dynamics
  of an epidemic due to population subdivision}}.
\newblock {\emph{\JournalTitle{Chaos}}} \textbf{\bibinfo{volume}{30}},
  \bibinfo{pages}{101102} (\bibinfo{year}{2020}).

\bibitem{das2021scaling}
\bibinfo{author}{Das, S.~K.}
\newblock \bibinfo{journal}{\bibinfo{title}{A scaling investigation of pattern
  in the spread of covid-19: universality in real data and a predictive
  analytical description}}.
\newblock {\emph{\JournalTitle{Proc. R. Soc. A.}}}
  \textbf{\bibinfo{volume}{477}}, \bibinfo{pages}{20200689}
  (\bibinfo{year}{2021}).

\bibitem{yaari2013modelling}
\bibinfo{author}{Yaari, R.}, \bibinfo{author}{Katriel, G.},
  \bibinfo{author}{Huppert, A.}, \bibinfo{author}{Axelsen, J.} \&
  \bibinfo{author}{Stone, L.}
\newblock \bibinfo{journal}{\bibinfo{title}{Modelling seasonal influenza: the
  role of weather and punctuated antigenic drift}}.
\newblock {\emph{\JournalTitle{J. R. Soc. Interface.}}}
  \textbf{\bibinfo{volume}{10}}, \bibinfo{pages}{20130298}
  (\bibinfo{year}{2013}).

\bibitem{zhao2021contagion}
\bibinfo{author}{Zhao, Y.}, \bibinfo{author}{Huepe, C.} \&
  \bibinfo{author}{Romanczuk, P.}
\newblock \bibinfo{journal}{\bibinfo{title}{Contagion dynamics in
  self-organized systems of self-propelled agents}}.
\newblock {\emph{\JournalTitle{arXiv preprint arXiv:2103.12618}}}
  (\bibinfo{year}{2021}).

\bibitem{norambuena2020understanding}
\bibinfo{author}{Norambuena, A.}, \bibinfo{author}{Valencia, F.~J.} \&
  \bibinfo{author}{Guzm{\'a}n-Lastra, F.}
\newblock \bibinfo{journal}{\bibinfo{title}{Understanding contagion dynamics
  through microscopic processes in active brownian particles}}.
\newblock {\emph{\JournalTitle{Scientific Reports}}}
  \textbf{\bibinfo{volume}{10}}, \bibinfo{pages}{20845} (\bibinfo{year}{2020}).

\bibitem{day2006insights}
\bibinfo{author}{Day, T.} \& \bibinfo{author}{Gandon, S.}
\newblock \bibinfo{journal}{\bibinfo{title}{Insights from price's equation into
  evolutionary epidemiology}}.
\newblock {\emph{\JournalTitle{Disease evolution: models, concepts, and data
  analyses}}} \textbf{\bibinfo{volume}{71}}, \bibinfo{pages}{23--44}
  (\bibinfo{year}{2006}).

\bibitem{day2007applying}
\bibinfo{author}{Day, T.} \& \bibinfo{author}{Gandon, S.}
\newblock \bibinfo{journal}{\bibinfo{title}{Applying population-genetic models
  in theoretical evolutionary epidemiology}}.
\newblock {\emph{\JournalTitle{Ecology Letters}}}
  \textbf{\bibinfo{volume}{10}}, \bibinfo{pages}{876--888}
  (\bibinfo{year}{2007}).

\bibitem{day2004general}
\bibinfo{author}{Day, T.} \& \bibinfo{author}{Proulx, S.~R.}
\newblock \bibinfo{journal}{\bibinfo{title}{A general theory for the
  evolutionary dynamics of virulence}}.
\newblock {\emph{\JournalTitle{The American Naturalist}}}
  \textbf{\bibinfo{volume}{163}}, \bibinfo{pages}{E40--E63}
  (\bibinfo{year}{2004}).

\bibitem{koelle2011dimensionless}
\bibinfo{author}{Koelle, K.}, \bibinfo{author}{Ratmann, O.},
  \bibinfo{author}{Rasmussen, D.~A.}, \bibinfo{author}{Pasour, V.} \&
  \bibinfo{author}{Mattingly, J.}
\newblock \bibinfo{journal}{\bibinfo{title}{A dimensionless number for
  understanding the evolutionary dynamics of antigenically variable rna
  viruses}}.
\newblock {\emph{\JournalTitle{Proceedings of the Royal Society B: Biological
  Sciences}}} \textbf{\bibinfo{volume}{278}}, \bibinfo{pages}{3723--3730}
  (\bibinfo{year}{2011}).

\bibitem{koelle2010two}
\bibinfo{author}{Koelle, K.}, \bibinfo{author}{Khatri, P.},
  \bibinfo{author}{Kamradt, M.} \& \bibinfo{author}{Kepler, T.~B.}
\newblock \bibinfo{journal}{\bibinfo{title}{A two-tiered model for simulating
  the ecological and evolutionary dynamics of rapidly evolving viruses, with an
  application to influenza}}.
\newblock {\emph{\JournalTitle{Journal of The Royal Society Interface}}}
  \textbf{\bibinfo{volume}{7}}, \bibinfo{pages}{1257--1274}
  (\bibinfo{year}{2010}).

\bibitem{he2013inferring}
\bibinfo{author}{He, D.}, \bibinfo{author}{Dushoff, J.}, \bibinfo{author}{Day,
  T.}, \bibinfo{author}{Ma, J.} \& \bibinfo{author}{Earn, D.~J.}
\newblock \bibinfo{journal}{\bibinfo{title}{Inferring the causes of the three
  waves of the 1918 influenza pandemic in england and wales}}.
\newblock {\emph{\JournalTitle{Proceedings of the Royal Society B: Biological
  Sciences}}} \textbf{\bibinfo{volume}{280}}, \bibinfo{pages}{20131345}
  (\bibinfo{year}{2013}).

\bibitem{boni2004influenza}
\bibinfo{author}{Boni, M.~F.}, \bibinfo{author}{Gog, J.~R.},
  \bibinfo{author}{Andreasen, V.} \& \bibinfo{author}{Christiansen, F.~B.}
\newblock \bibinfo{journal}{\bibinfo{title}{Influenza drift and epidemic size:
  the race between generating and escaping immunity}}.
\newblock {\emph{\JournalTitle{Theoretical population biology}}}
  \textbf{\bibinfo{volume}{65}}, \bibinfo{pages}{179--191}
  (\bibinfo{year}{2004}).

\bibitem{levin1981selection}
\bibinfo{author}{Levin, S.} \& \bibinfo{author}{Pimentel, D.}
\newblock \bibinfo{journal}{\bibinfo{title}{Selection of intermediate rates of
  increase in parasite-host systems}}.
\newblock {\emph{\JournalTitle{The American Naturalist}}}
  \textbf{\bibinfo{volume}{117}}, \bibinfo{pages}{308--315}
  (\bibinfo{year}{1981}).

\bibitem{maier2020effective}
\bibinfo{author}{Maier, B.~F.} \& \bibinfo{author}{Brockmann, D.}
\newblock \bibinfo{journal}{\bibinfo{title}{Effective containment explains
  subexponential growth in recent confirmed covid-19 cases in china}}.
\newblock {\emph{\JournalTitle{Science}}} \textbf{\bibinfo{volume}{368}},
  \bibinfo{pages}{742--746} (\bibinfo{year}{2020}).

\bibitem{dehning2020inferring}
\bibinfo{author}{Dehning, J.} \emph{et~al.}
\newblock \bibinfo{journal}{\bibinfo{title}{Inferring change points in the
  spread of covid-19 reveals the effectiveness of interventions}}.
\newblock {\emph{\JournalTitle{Science}}} \textbf{\bibinfo{volume}{369}}
  (\bibinfo{year}{2020}).

\bibitem{te2020effects}
\bibinfo{author}{Te~Vrugt, M.}, \bibinfo{author}{Bickmann, J.} \&
  \bibinfo{author}{Wittkowski, R.}
\newblock \bibinfo{journal}{\bibinfo{title}{Effects of social distancing and
  isolation on epidemic spreading modeled via dynamical density functional
  theory}}.
\newblock {\emph{\JournalTitle{Nat. Commun.}}} \textbf{\bibinfo{volume}{11}},
  \bibinfo{pages}{5576} (\bibinfo{year}{2020}).

\bibitem{duran2021more}
\bibinfo{author}{Duran-Olivencia, M.~A.} \& \bibinfo{author}{Kalliadasis, S.}
\newblock \bibinfo{journal}{\bibinfo{title}{More than a year after the onset of
  the covid-19 pandemic in the uk: lessons learned from a minimalistic model
  capturing essential features including social awareness and policy making}}.
\newblock {\emph{\JournalTitle{medRxiv}}}  (\bibinfo{year}{2021}).

\bibitem{te2021containing}
\bibinfo{author}{te~Vrugt, M.}, \bibinfo{author}{Bickmann, J.} \&
  \bibinfo{author}{Wittkowski, R.}
\newblock \bibinfo{journal}{\bibinfo{title}{Containing a pandemic:
  nonpharmaceutical interventions and the "second wave"}}.
\newblock {\emph{\JournalTitle{J. Phys. Commun.}}}
  \textbf{\bibinfo{volume}{5}}, \bibinfo{pages}{055008} (\bibinfo{year}{2021}).

\bibitem{lasser2021assessing}
\bibinfo{author}{Lasser, J.} \emph{et~al.}
\newblock \bibinfo{journal}{\bibinfo{title}{Assessing the impact of sars-cov-2
  prevention measures in schools by means of agent-based simulations calibrated
  to cluster tracing data}}.
\newblock {\emph{\JournalTitle{medRxiv}}}  (\bibinfo{year}{2021}).

\bibitem{desvars2020structured}
\bibinfo{author}{Desvars-Larrive, A.} \emph{et~al.}
\newblock \bibinfo{journal}{\bibinfo{title}{A structured open dataset of
  government interventions in response to covid-19}}.
\newblock {\emph{\JournalTitle{Sci Data}}} \textbf{\bibinfo{volume}{7}},
  \bibinfo{pages}{285} (\bibinfo{year}{2020}).

\bibitem{bittihn2021local}
\bibinfo{author}{Bittihn, P.}, \bibinfo{author}{Hupe, L.},
  \bibinfo{author}{Isensee, J.} \& \bibinfo{author}{Golestanian, R.}
\newblock \bibinfo{journal}{\bibinfo{title}{Local measures enable covid-19
  containment with fewer restrictions due to cooperative effects}}.
\newblock {\emph{\JournalTitle{EClinicalMedicine}}}
  \textbf{\bibinfo{volume}{32}}, \bibinfo{pages}{100718}
  (\bibinfo{year}{2021}).

\bibitem{zhang2021epidemic}
\bibinfo{author}{Zhang, X.} \emph{et~al.}
\newblock \bibinfo{journal}{\bibinfo{title}{Epidemic spreading under pathogen
  evolution}}.
\newblock {\emph{\JournalTitle{arXiv preprint arXiv:2102.11066}}}
  (\bibinfo{year}{2021}).

\bibitem{contreras2021challenges}
\bibinfo{author}{Contreras, S.} \emph{et~al.}
\newblock \bibinfo{journal}{\bibinfo{title}{The challenges of containing
  sars-cov-2 via test-trace-and-isolate}}.
\newblock {\emph{\JournalTitle{Nat. Commun.}}} \textbf{\bibinfo{volume}{12}},
  \bibinfo{pages}{1--13} (\bibinfo{year}{2021}).

\bibitem{estrada2020covid}
\bibinfo{author}{Estrada, E.}
\newblock \bibinfo{journal}{\bibinfo{title}{Covid-19 and sars-cov-2. modeling
  the present, looking at the future}}.
\newblock {\emph{\JournalTitle{Physics Reports}}}
  \textbf{\bibinfo{volume}{869}}, \bibinfo{pages}{1--51}
  (\bibinfo{year}{2020}).

\bibitem{yang2020mathematical}
\bibinfo{author}{Yang, C.} \& \bibinfo{author}{Wang, J.}
\newblock \bibinfo{journal}{\bibinfo{title}{A mathematical model for the novel
  coronavirus epidemic in wuhan, china}}.
\newblock {\emph{\JournalTitle{Mathematical biosciences and engineering: MBE}}}
  \textbf{\bibinfo{volume}{17}}, \bibinfo{pages}{2708} (\bibinfo{year}{2020}).

\bibitem{gonzalez2021impact}
\bibinfo{author}{Gonzalez-Parra, G.},
  \bibinfo{author}{Mart{\'\i}nez-Rodr{\'\i}guez, D.} \&
  \bibinfo{author}{Villanueva-Mic{\'o}, R.~J.}
\newblock \bibinfo{journal}{\bibinfo{title}{Impact of a new sars-cov-2 variant
  on the population: A mathematical modeling approach}}.
\newblock {\emph{\JournalTitle{Mathematical and Computational Applications}}}
  \textbf{\bibinfo{volume}{26}}, \bibinfo{pages}{25} (\bibinfo{year}{2021}).

\bibitem{fudolig2020local}
\bibinfo{author}{Fudolig, M.} \& \bibinfo{author}{Howard, R.}
\newblock \bibinfo{journal}{\bibinfo{title}{The local stability of a modified
  multi-strain sir model for emerging viral strains}}.
\newblock {\emph{\JournalTitle{PloS one}}} \textbf{\bibinfo{volume}{15}},
  \bibinfo{pages}{e0243408} (\bibinfo{year}{2020}).

\bibitem{tian2021harnessing}
\bibinfo{author}{Tian, L.} \emph{et~al.}
\newblock \bibinfo{journal}{\bibinfo{title}{Harnessing peak transmission around
  symptom onset for non-pharmaceutical intervention and containment of the
  covid-19 pandemic}}.
\newblock {\emph{\JournalTitle{Nat. Commun.}}} \textbf{\bibinfo{volume}{12}},
  \bibinfo{pages}{1--12} (\bibinfo{year}{2021}).

\bibitem{meidan2021alternating}
\bibinfo{author}{Meidan, D.} \emph{et~al.}
\newblock \bibinfo{journal}{\bibinfo{title}{Alternating quarantine for
  sustainable epidemic mitigation}}.
\newblock {\emph{\JournalTitle{Nat. Commun.}}} \textbf{\bibinfo{volume}{12}},
  \bibinfo{pages}{1--12} (\bibinfo{year}{2021}).

\bibitem{hasan2021emergence}
\bibinfo{author}{Hasan, M.~M.} \emph{et~al.}
\newblock \bibinfo{journal}{\bibinfo{title}{Emergence of highly infectious
  sars-cov-2 variants in bangladesh: the need for systematic genetic
  surveillance as a public health strategy}}.
\newblock {\emph{\JournalTitle{Tropical Medicine and Health}}}
  \textbf{\bibinfo{volume}{49}}, \bibinfo{pages}{1--3} (\bibinfo{year}{2021}).

\bibitem{essar2021covid}
\bibinfo{author}{Essar, M.~Y.} \emph{et~al.}
\newblock \bibinfo{journal}{\bibinfo{title}{Covid-19 and multiple crises in
  afghanistan: an urgent battle}}.
\newblock {\emph{\JournalTitle{Conflict and Health}}}
  \textbf{\bibinfo{volume}{15}}, \bibinfo{pages}{1--3} (\bibinfo{year}{2021}).

\bibitem{islam2022food}
\bibinfo{author}{Islam, Z.} \emph{et~al.}
\newblock \bibinfo{journal}{\bibinfo{title}{Food security, conflict, and
  covid-19: perspective from afghanistan}}.
\newblock {\emph{\JournalTitle{The American journal of tropical medicine and
  hygiene}}} \textbf{\bibinfo{volume}{106}}, \bibinfo{pages}{21}
  (\bibinfo{year}{2022}).

\bibitem{aborode2021impact}
\bibinfo{author}{Aborode, A.~T.} \emph{et~al.}
\newblock \bibinfo{journal}{\bibinfo{title}{Impact of poor disease surveillance
  system on covid-19 response in africa: time to rethink and rebuilt}}.
\newblock {\emph{\JournalTitle{Clinical epidemiology and global health}}}
  \textbf{\bibinfo{volume}{12}}, \bibinfo{pages}{100841}
  (\bibinfo{year}{2021}).

\bibitem{aborode2021dengue}
\bibinfo{author}{Aborode, A.~T.} \emph{et~al.}
\newblock \bibinfo{journal}{\bibinfo{title}{Dengue and coronavirus disease
  (covid-19) syndemic: Double threat to an overburdened healthcare system in
  africa}}.
\newblock {\emph{\JournalTitle{The International Journal of Health Planning and
  Management}}}  (\bibinfo{year}{2021}).

\bibitem{rackimuthu2021covid}
\bibinfo{author}{Rackimuthu, S.}, \bibinfo{author}{Hasan, M.~M.},
  \bibinfo{author}{Bardhan, M.} \& \bibinfo{author}{Essar, M.~Y.}
\newblock \bibinfo{journal}{\bibinfo{title}{Covid-19 vaccination strategies and
  policies in india: the need for further re-evaluation is a pressing
  priority}}.
\newblock {\emph{\JournalTitle{The International Journal of Health Planning and
  Management}}}  (\bibinfo{year}{2021}).

\bibitem{o2013chaos}
\bibinfo{author}{O’Regan, S.~M.} \emph{et~al.}
\newblock \bibinfo{journal}{\bibinfo{title}{Chaos in a seasonally perturbed sir
  model: avian influenza in a seabird colony as a paradigm}}.
\newblock {\emph{\JournalTitle{Journal of mathematical biology}}}
  \textbf{\bibinfo{volume}{67}}, \bibinfo{pages}{293--327}
  (\bibinfo{year}{2013}).

\bibitem{liu2020reproductive}
\bibinfo{author}{Liu, Y.}, \bibinfo{author}{Gayle, A.~A.},
  \bibinfo{author}{Wilder-Smith, A.} \& \bibinfo{author}{Rocklöv, J.}
\newblock \bibinfo{journal}{\bibinfo{title}{{The reproductive number of
  COVID-19 is higher compared to SARS coronavirus}}}.
\newblock {\emph{\JournalTitle{Journal of Travel Medicine}}}
  \textbf{\bibinfo{volume}{27}}, \doiprefix\url{10.1093/jtm/taaa021}
  (\bibinfo{year}{2020}).

\bibitem{assessment2021risk}
\bibinfo{author}{Assessment, R.~R.}
\newblock \bibinfo{journal}{\bibinfo{title}{Risk related to the spread of new
  sars-cov-2 variants of concern in the eu/eea--first update}}.
\newblock {\emph{\JournalTitle{European Centre for Disease Prevention and
  Control An agency of the European Union}}}  (\bibinfo{year}{2021}).

\bibitem{campbell2021increased}
\bibinfo{author}{Campbell, F.} \emph{et~al.}
\newblock \bibinfo{journal}{\bibinfo{title}{Increased transmissibility and
  global spread of sars-cov-2 variants of concern as at june 2021}}.
\newblock {\emph{\JournalTitle{Eurosurveillance}}}
  \textbf{\bibinfo{volume}{26}}, \bibinfo{pages}{2100509}
  (\bibinfo{year}{2021}).

\bibitem{nishiura2021relative}
\bibinfo{author}{Nishiura, H.} \emph{et~al.}
\newblock \bibinfo{title}{Relative reproduction number of sars-cov-2 omicron
  (b. 1.1. 529) compared with delta variant in south africa}
  (\bibinfo{year}{2021}).

\end{thebibliography}

\providecommand{\noopsort}[1]{}\providecommand{\singleletter}[1]{#1}%

\end{document}